\documentclass[aps, prd, 10pt, twocolumn, superscriptaddress,noshowpacs, preprintnumbers, longbibliography,nofootinbib,bibnotes,floatfix]{revtex4-1}
\usepackage[T1]{fontenc}

\usepackage{amsmath}
\usepackage{amsfonts}
\usepackage{amssymb}
\usepackage{bbold}
\usepackage{epsfig}
\usepackage{graphicx}
\usepackage[caption=false]{subfig} 
\usepackage{bm}
\usepackage{array}
\usepackage{hyperref}
\usepackage{listings}
\usepackage{color}
\usepackage{float}
\usepackage{url} 
\usepackage[normalem]{ulem}
\usepackage{breqn}
\usepackage{makecell}

\usepackage{lineno}
\usepackage{lipsum}
\usepackage{physics}
\usepackage{mathtools}

\usepackage{tabularx}
\usepackage[export]{adjustbox}
\usepackage{multirow}
\usepackage{citesort}
\usepackage{soul}
\usepackage{bm}
\usepackage{xcolor}
\usepackage[utf8]{inputenc}
\usepackage[normalem]{ulem}
\usepackage{accents}
\usepackage{tikz}

\definecolor{Wildstrawberry}{rgb}{1.0, 0.26, 0.64}

\usepackage{tikz,xcolor,hyperref}

\definecolor{lime}{HTML}{A6CE39}
\DeclareRobustCommand{\orcidicon}{\hspace{-1mm}
	\begin{tikzpicture}
	\draw[lime, fill=lime] (0,0) 
	circle [radius=0.16] 
	node[white] {{\fontfamily{qag}\selectfont \tiny \,ID}};
	\draw[white, fill=white] (-0.0525,0.095) 
	circle [radius=0.007];
	\end{tikzpicture}
	\hspace{-3mm}
}

\foreach \x in {A, ..., Z}{\expandafter\xdef\csname orcid\x\endcsname{\noexpand\href{https://orcid.org/\csname orcidauthor\x\endcsname}
			{\noexpand\orcidicon}}
}

\newcommand{\affiliationA}{Universidade Estadual de Campinas, Instituto de F\'{\i}sica Gleb Wataghin, R. S\'ergio Buarque de Holanda, 777, Brazil} %
\newcommand{\affiliationB}{Instituto de F\'isica, Universidade de S\~ao Paulo, C.P. 66.318, 05315-970 S\~ao Paulo, Brazil} %

\begin{document}

\title{Time-of-Flight Constraints on Neutrino Millicharge from Supernova Neutrinos in Galactic Magnetic Fields}

\author{Pedro Dedin Neto\orcidA{}}
\affiliation{\affiliationB}
\author{AmirFarzan Esmaeili\orcidB{}}
\affiliation{\affiliationA}
\author{Guilherme A. Nogueira\orcidC{}}
\affiliation{\affiliationA}
\author{Pedro Cunha de Holanda\orcidD{}}
\affiliation{\affiliationA}
\author{Ernesto Kemp\orcidE{}}
\affiliation{\affiliationA}

\newcommand{\PD}[1]{{\color{red}\textbf{P. Dedin:} #1}}
\newcommand{\PCH}[1]{{\color{blue}\textbf{P. Holanda:} #1}}
\newcommand{\AFE}[1]{{\color{teal}\textbf{Amir:} #1}}
\newcommand{\GN}[1]{\textcolor{Wildstrawberry}{\textbf{G. Nogueira:} { #1}}}
\newcommand{\gn}[1]{\textcolor{Wildstrawberry}{{ #1}}}
\newcommand{\EK}[1]{{\color{purple}\textbf{E. Kemp:} #1}}

\date{\today}

\begin{abstract}

A millicharged neutrino propagating through magnetic fields experiences a small Lorentz-force deflection, which induces a geometric time delay. In the ultra-relativistic regime relevant for supernova neutrinos, this delay scales as $q_\nu^2 E_\nu^{-2}$, where $q_\nu$ and $E_\nu$ denote the neutrino millicharge and energy, respectively, and thus shares the same leading energy dependence as the standard time-of-flight delay induced by neutrino mass. Motivated by this similarity, we propose a framework to reinterpret supernova time-of-flight limits on neutrino mass as constraints on neutrino millicharge. We express both effects in terms of a common $E_\nu^{-2}$ dispersion coefficient and compute the millicharge-induced contribution using a line-of-sight-dependent magnetic delay kernel, extending the original SN1987A uniform-field estimate. Applying this translation to existing SN1987A limits and to projected sensitivities for future Galactic core-collapse supernova observations, we obtain bounds ranging from the $\sim 10^{-17}\, e$ level for SN1987A to the low-$10^{-19}\, e$ regime for next-generation Galactic bursts, with optimistic combinations of detector sensitivity and Galactic sightline approaching $\sim 10^{-20}\, e$. We compare these results with other bounds in the literature and discuss how nonzero neutrino mass affects the interpretation.

\end{abstract}

\maketitle
\section{Introduction}
\label{sec:introduction}

In the minimal Standard Model (SM) with the usual hypercharge assignments, neutrinos are electrically neutral. More generally, electric charge is associated with the generator \(Q\) of the unbroken electromagnetic symmetry and, in the electroweak theory, is given by $Q = T_3 + \frac{Y}{2}$, where \(T_3\) is the third weak-isospin generator and \(Y\) is the weak hypercharge. The electric charges of quarks and leptons are therefore fixed by their electroweak quantum numbers. Whether this charge assignment is uniquely enforced by the theory is tied to the more general question of electric-charge quantization.

From a theoretical perspective, exact electric charge quantization is not automatic in all extensions of the SM. Although anomaly cancellation and the Yukawa structure strongly constrain the allowed hypercharge assignments, it has long been recognized that more general Abelian gauge structures can permit electric charge dequantization \cite{PhysRevD.41.271,Foot:1992ui,1990MPLA....5.2721F,PhysRevD.44.2118}. In modern terms, one may view this as replacing the standard hypercharge embedding by a linear combination involving an additional anomaly-free $U(1)_X$ symmetry under which neutrinos transform nontrivially. 

Neutrino oscillation experiments have established that neutrinos are massive and mixed, demonstrating that the minimal SM is incomplete and that new physics is required beyond its original formulation \cite{Super-Kamiokande:1998kpq,SNO:2002tuh,KamLAND:2002uet}. Although neutrino mass by itself does not imply a nonzero neutrino electric charge, the extensions introduced to account for neutrino masses can modify the conditions under which charge quantization is realized. In viable millicharged-neutrino scenarios, this issue is closely tied to the Dirac nature of neutrinos: once neutrinos carry electric charge, Majorana mass terms are forbidden by electromagnetic gauge invariance, so charged neutrinos are necessarily Dirac fermions \cite{Giunti:2014ixa, Jana:2025eii}. Moreover, recent analyses indicate that not all anomaly-free \(U(1)_X\) realizations remain phenomenologically viable once the observed neutrino masses and mixing are imposed: minimal flavor-dependent symmetries such as \(U(1)_{L_\alpha-L_\beta}\) are disfavored, whereas flavor-universal possibilities such as \(U(1)_{B-L}\) or \(U(1)_L\) remain viable candidates for accommodating tiny neutrino charges~\cite{Jana:2025eii}. More generally, minimal Dirac extensions with right-handed neutrinos can permit electric charge dequantization through anomaly-free Abelian symmetries such as $B-L$ \cite{Foot:1992ui, PhysRevD.41.271, Jana:2025eii}.

In this context, a nonzero electric charge is especially interesting among the possible electromagnetic properties of neutrinos, because it is directly tied to the problem of charge quantization and to possible extensions of the SM gauge structure. More broadly, neutrino electromagnetic properties provide a window on physics beyond the Standard Model, including the Dirac or Majorana nature of neutrinos and the possible existence of hidden sectors or additional gauge interactions \cite{1982NuPhB.206..359S,BrogginiGiuntiStudenikin2012,Giunti:2014ixa,Giunti:2024gec}. In particular, a neutrino millicharge can arise in representative beyond the Standard Model (BSM) scenarios either through electric charge dequantization or through kinetic mixing involving an additional Abelian gauge sector \cite{Foot:1992ui,Holdom:1985ag}.

Existing bounds on the neutrino millicharge have been derived from a broad range of laboratory, astrophysical, and cosmological probes, including reactor neutrino scattering, stellar cooling, big bang nucleosynthesis, and electromagnetic effects in astrophysical environments \cite{Raffelt:1999gv, 2000JHEP...05..003D, Giunti:2014ixa, Chen:2014dsa, AtzoriCorona:2022jeb, Khan:2022bel, Brudanin:2014iya, Das:2020egb}. Although these constraints already require the neutrino charge to be extremely small, they probe distinct physical effects and do not exhaust the range of possible observables. In particular, the propagation of neutrinos over astrophysical distances through magnetic fields offers a complementary way to test tiny electric charges through the cumulative effect of small Lorentz force deflections.

Core-collapse supernovae (CCSNe) are especially well suited to this purpose. They emit intense bursts of MeV neutrinos over timescales of order seconds, and characteristic temporal features of the signal---such as the neutronization burst or, in failed explosions, an abrupt cutoff associated with black hole formation ---can provide sharp timing markers for time-of-flight (ToF) studies. Even tiny energy dependent propagation delays can therefore accumulate over astrophysical baselines and leave an observable imprint on the detected burst profile. 

In particular, the idea that neutrino mass could be constrained from the arrival time delay of neutrinos from a distant transient source goes back to Zatsepin, who pointed out that a massive neutrino acquires a ToF delay scaling as \(m_\nu^2 E_\nu^{-2}\) over a fixed baseline \cite{Zatsepin:1968kt}. This basic logic underlies the long standing supernova ToF literature, from early discussions and reanalyses of the SN1987A signal \cite{Schramm:1989rf, Loredo:2001rx, Pagliaroli:2010ik} to modern forecasts for a future Galactic burst in detectors such as DUNE, Hyper Kamiokande, JUNO and IceCube \cite{Beacom:2000qy, Nardi:2004zg, Pompa:2022cxc, Parker:2023cos, Denton:2024mlb, Ellis:2012ji}. 

The same timing logic can be applied to a millicharged neutrino, although the microscopic origin of the delay is different. A neutrino carrying electric charge \(q_\nu\) is deflected by magnetic fields during propagation, and the associated increase in path length induces an additional geometric time delay. In the ultra relativistic regime relevant for the supernova neutrinos, this delay also scales as \(q_\nu^2 E_\nu^{-2}\) at leading order. Shortly after SN1987A, Barbiellini and Cocconi used this idea to derive a bound on the neutrino electric charge assuming propagation in a uniform magnetic field \cite{Barbiellini:1987zz}. Their estimate captured the basic physical effect, but both the magnetic field description and the statistical treatment of the supernova signal were simplified relative to the modern supernova ToF analyses.

In this work, we revisit the supernova constraint on neutrino millicharge within a modern ToF framework. The key point is that, at the phenomenological level, supernova timing analyses are sensitive to the coefficient of an energy dependent dispersion term in the detected burst profile. Since the standard mass induced delay and the magnetic delay of a millicharged neutrino share the same leading \(E_\nu^{-2}\) dependence, published and projected ToF bounds on neutrino mass can be systematically reinterpreted as bounds on neutrino millicharge. We implement this program by expressing both effects in terms of a common dispersion coefficient and by evaluating the millicharge induced contribution with a line-of-sight dependent kernel determined by the Galactic magnetic field model, thereby extending the original uniform field estimate of Ref. \cite{Barbiellini:1987zz}. The resulting translated sensitivity depends explicitly on both the detector level ToF reach and the astrophysical propagation along the source direction, so the inferred charge bound is not a single universal number but a line-of-sight dependent sensitivity. Note that the present translation is intended for time-of-flight analyses that constrain a single effective leading $E_\nu^{-2}$ dispersion coefficient for the detected signal; analyses that become sensitive to the individual neutrino mass eigenstates fall outside this minimal recast.

The paper is organized as follows. In Sec.~\ref{sec:formalism}, we introduce the ToF framework and define the dispersion coefficient parametrization used throughout the paper. In Sec.~\ref{sec:results}, we translate the relevant supernova ToF bounds into limits on neutrino millicharge using our magnetic delay formalism. In Sec.~\ref{sec:discussion}, we compare our results with other bounds in the literature, discuss their interpretation and main caveats, and present our conclusions.

\section{Framework}
\label{sec:formalism}

Time-of-flight (ToF) studies of supernova neutrinos probe energy dependent distortions of the observed burst time structure. Existing ToF results are commonly quoted as constraints on an effective neutrino mass, corresponding to the usual interpretation in which the relevant propagation induced delay is attributed entirely to neutrino mass. 

The recast developed in this work exploits the fact that the magnetic field induced delay of a millicharged neutrino has the same leading energy dependence. This allows a published bound on neutrino mass to be reinterpreted as a bound on neutrino millicharge at the level of the leading propagation induced delay.

\subsection{Dispersion coefficient translation}
\label{subsec:translation}

For an ultra relativistic neutrino of energy $E$ and mass $m_\nu$, the corresponding delay over a source distance $L$ is
\begin{equation}
\Delta t_m(E)
\simeq 
\frac{L}{2c}\,\frac{m_\nu^2}{E^2}\, .
\label{eq:dt_mass}
\end{equation}
It is therefore useful to isolate the coefficient of this common $E^{-2}$ scaling and write the leading propagation induced delay as
\begin{equation}
\Delta t_{\rm prop}(E)\simeq \frac{A_{\rm ToF}}{E^2}\, ,
\label{eq:A2_def}
\end{equation}
where $A_{\rm ToF}$ denotes the \textit{dispersion coefficient} of the leading delay term. In the mass-only interpretation, this coefficient reduces to
\begin{equation}
A_{\rm ToF}=A_m \equiv \frac{L}{2c}\,m_\nu^2\, .
\label{eq:Am_def}
\end{equation}

At leading order, the timing sensitivity of a ToF analysis can thus be expressed as sensitivity to the coefficient of the leading propagation delay. The usual interpretation in terms of neutrino mass corresponds to one particular realization of this coefficient. More generally, any mechanism with the same leading energy dependence contributes to the same effective quantity, so that
\begin{equation}
A_{\rm ToF} \equiv 
A_m(m_\nu^2)+A_q(q_\nu^2)+A_{\rm other}\, ,
\label{eq:A_decomposition}
\end{equation}
where $A_m$ and $A_q$ are the dispersion coefficients associated with the mass-induced and millicharge-induced delays, respectively, and $A_{\rm other}$ collects any additional contribution whose leading ToF effect scales as $E^{-2}$ energy dependence.

A reported upper bound $m_\nu<m_{\rm lim}$ therefore implies, under the mass-only interpretation,
\begin{equation}
A_{\rm ToF}<A_{\rm lim}\equiv\frac{L}{2c}\,m_{\rm lim}^2\, ,
\label{eq:Alim_from_mlim}
\end{equation}
i.e. for $A_q=A_{\rm other}=0$. Equation~\eqref{eq:Alim_from_mlim} maps a published ToF mass limit onto the more general coefficient entering our recast. Assuming instead that the entire constrained leading contribution is due to a millicharge, we require
\begin{equation}
A_q(q_\nu^2)\le A_{\rm lim}\, .
\label{eq:Aq_le_Alim}
\end{equation}

The problem is then reduced to computing the millicharge induced coefficient \(A_q\) for a specified magnetic field configuration.

\subsection{Millicharged neutrino time delay in magnetic fields}
\label{subsec:Bfield_timedelay}

A neutrino carrying electric charge $q_\nu$ is deflected by magnetic fields through the Lorentz force. In the small deflection regime relevant here, the resulting increase in path length produces a geometric arrival time delay proportional to $(q_\nu/E)^2$. The derivation of this scaling, together with the fixed endpoint kernel formulation used below, is given in Appendix~\ref{app:kernel_derivation}. In the parameter region relevant for the translated bounds obtained in this work, the net deflection remains extremely small, so the small angle approximation is well justified.

For Galactic supernova neutrinos, the magnetic delay is dominated by propagation through the Galactic magnetic field (GMF). In our numerical implementation, we model the GMF using the Jansson-Farrar (JF12) framework \cite{2012ApJ...757...14J}, with the corresponding parameter uncertainties propagated phenomenologically as described in Appendix~\ref{app:gmf}. The JF12 model explicitly includes regular, turbulent, and striated components, which we write as
\begin{equation}
\mathbf{B}=\mathbf{B}_{\rm reg}+\mathbf{B}_{\rm turb}+\mathbf{B}_{\rm str}\, ,
\label{eq:B_decomposition}
\end{equation}
where $\mathbf{B}_{\rm turb}$ is an isotropic random component with vanishing mean and finite correlation scale, while $\mathbf{B}_{\rm str}$ is a striated component, locally aligned with $\mathbf{B}_{\rm reg}$ but fluctuating in sign so that $\langle \mathbf{B}_{\rm str}\rangle=0$ while $\langle B_{\rm str}^2\rangle\neq 0$. We first consider the contribution of the regular component.

Let $\mathbf{B}_\perp(x)$ denote the magnetic field transverse to the unperturbed line-of-sight at position $x$ along the trajectory. In the small-angle limit, the delay induced by the regular GMF component can be written as
\begin{subequations}
\begin{align}
\Delta t_q^{\rm reg}(E)
&=
\frac{1}{2c}\left(\frac{q_\nu}{E}\right)^2
\notag\\
&\quad\times
\int_0^{L}\!{\rm d}x\!\int_0^{L}\!{\rm d}x'\,
\mathbf{B}_\perp(x)\cdot\mathbf{B}_\perp(x')\,K(x,x')\, ,
\label{eq:dtq_reg_kernel}
\\
K(x,x') &= \min(x,x')-\frac{xx'}{L}\, .
\label{eq:bb_kernel}
\end{align}
\end{subequations}
Here, the superscript ``reg'' indicates that only the regular GMF component is included. The kernel $K(x,x')$ is the geometric weighting appropriate to small transverse excursions with fixed source and observer positions.\footnote{Equivalently, $K(x,x')$ is the Green's function of the associated fixed-endpoint transverse-displacement problem; after rescaling to the unit interval, it is mathematically identical to the covariance kernel of a Brownian bridge. See Appendix~\ref{app:kernel_derivation}.}

It is convenient to define an effective field strength $B_{\rm eff}$ by
\begin{equation}
B_{\mathrm{eff}}^2
\equiv
\frac{12}{L^3}
\int_0^{L}\!{\rm d}x\!\int_0^{L}\!{\rm d}x'\,
\mathbf{B}_\perp(x)\cdot\mathbf{B}_\perp(x')\,K(x,x')\, .
\label{eq:Beff_def}
\end{equation}
With this definition, Eq.~\eqref{eq:dtq_reg_kernel} becomes
\begin{equation}
\Delta t_q^{\rm reg}(E)
=
\frac{L^3}{24c}\left(\frac{q_\nu\,B_{\rm eff}}{E}\right)^2
\equiv
\frac{A_q^{\rm reg}}{E^2}\, ,
\label{eq:dtq_reg_final}
\end{equation}
so that the corresponding dispersion coefficient is
\begin{equation}
A_q^{\rm reg}
=
\frac{L^3}{24c}\,q_\nu^{\,2}\,B_{\rm eff}^{\,2}\, .
\label{eq:Aq_reg}
\end{equation}

The stochastic GMF components may be treated in the regime where the propagation distance is much larger than their correlation scale. In that limit, the neutrino traverses many approximately independent magnetic domains along the line-of-sight, the accumulated transverse deflection behaves as a random walk, and the resulting time delay is well described by its ensemble average over realizations of the stochastic field. In this regime, the delay is controlled by the field variance and by the integral scale of the transverse two-point correlation function, $\lambda_B$, as defined in Appendix~\ref{app:kernel_derivation}.

Using the stochastic field limit of the kernel formulation, the turbulent contribution takes the form
\begin{equation}
\left\langle \Delta t_q^{\rm turb}(E)\right\rangle
=
\frac{L^2\lambda_B^{\rm turb}}{9c}
\left(\frac{q_\nu B_{\rm rms}^{\rm turb}}{E}\right)^2
\equiv
\frac{A_q^{\rm turb}}{E^2}\, ,
\label{eq:dtq_turb}
\end{equation}
so that
\begin{equation}
A_q^{\rm turb}
=
\frac{q_\nu^2 L^2\lambda_B^{\rm turb}}{9c}
\left(B_{\rm rms}^{\rm turb}\right)^2\, .
\label{eq:Aq_turb}
\end{equation}
Here $B_{\rm rms}^{\rm turb}$ is the full rms turbulent field strength; for an isotropic random field, the corresponding transverse variance satisfies $\langle |\mathbf{B}_{{\rm turb},\perp}|^2\rangle_x = (2/3)(B_{\rm rms}^{\rm turb})^2$, where $\langle\rangle_x$ denotes the average along the line-of-sight.

For the striated component, the general stochastic kernel expression remains valid, but the isotropic relation between the transverse variance and the full three dimensional rms field no longer applies. It is therefore convenient to parameterize the result directly in terms of the transverse variance
\begin{equation}
\mathcal{B}_{\rm str}^2
\equiv
\left\langle |\mathbf{B}_{{\rm str},\perp}|^2 \right\rangle_x\, ,
\label{eq:Bstr_perp_var}
\end{equation}

\noindent Thus, the corresponding dispersion coefficient is 
\begin{equation}
A_q^{\rm str}
=
\frac{q_\nu^2 L^2\lambda_B^{\rm str}}{6c}\,
\mathcal{B}_{\rm str}^2\, .
\label{eq:Aq_str}
\end{equation}
In practice, for each line-of-sight we generate multiple independent realizations of the JF12 turbulent and striated fields and estimate the corresponding combinations $(B_{\rm rms}^{\rm turb})^2\lambda_B^{\rm turb}$ and $\mathcal{B}_{\rm str}^2\lambda_B^{\rm str}$ directly from the sampled transverse field components; see Appendix~\ref{app:gmf} for details. Their percentiles are then used to propagate the stochastic field uncertainty into the corresponding contributions to the time delay kernel.

The total millicharge contribution to the leading delay coefficient is therefore the sum of the regular and stochastic pieces,
\begin{equation}
A_q = A_q^{\rm reg}+A_q^{\rm turb}+A_q^{\rm str}\, .
\label{eq:Aq_total}
\end{equation}

Finally, the contribution of extragalactic magnetic fields (EGMFs) is negligible for the Galactic supernova distances relevant here. Rotation measure studies constrain EGMFs to be at most at the nG level on Mpc coherence scales, and often much smaller in low density environments~\cite{Pshirkov:2015tua}. Since a coherent contribution scales as $B^2L^3$, even an optimistic estimate with $B_{\rm EGMF}\lesssim 1~{\rm nG}$ over a path of a few$\times 10$~kpc is at least $\sim 10^{-4}$ below a typical Galactic contribution with $B_{\rm GMF}\sim \mu{\rm G}$ and $L_{\rm gal}\sim 10$~kpc. For weaker void-like extragalactic fields, the suppression is far stronger. We therefore neglect the EGMF contribution throughout.
\subsection{From published ToF limits to millicharge constraints}
\label{subsec:translation_recipe}


\begin{table*}[t]
\centering
\caption{
Published and projected supernova ToF inputs used in this work. The quoted mass limit $m_{\rm lim}$, confidence level (C.L.), and assumed source distance $L$ are taken from the cited references. The corresponding upper limit on the leading $E^{-2}$ dispersion coefficient is obtained from Eq.~\eqref{eq:Alim_from_mlim}.
For DUNE, we list optimistic and conservative entries corresponding to the lower and upper endpoints of the mass sensitivity ranges quoted in Refs.~\cite{Pompa:2022cxc}.
}
\label{tab:tof_inputs}
\renewcommand{\arraystretch}{1.35}
\begin{tabular}{lcccc}
\hline\hline
\textbf{Analysis (Ref.)} &
\textbf{$m_{\rm lim}$ [eV]} &
\textbf{C.L.} &
\textbf{$L$} &
\textbf{$A_{\rm lim}$ [s\,MeV$^2$]} \\
\hline
SN1987A (model indep.)~\cite{Schramm:1989rf} &
$\lesssim 30$ &
(order of mag.) &
50 kpc &
$\lesssim 2.32\times 10^{3}$ \\
SN1987A (model dep.)~\cite{Loredo:2001rx} &
$<5.7$ &
95\% &
50 kpc &
$<8.36\times 10^{1}$ \\
\hline
Galactic CCSN (SK)~\cite{Pagliaroli:2010ik} &
$<0.8$ &
95\% &
10 kpc &
$<3.29\times 10^{-1}$ \\
Galactic CCSN (IceCube)~\cite{Ellis:2012ji} &
$<0.14$ &
95\% &
10 kpc &
$<1.01\times 10^{-2}$ \\
Galactic CCSN (DUNE, optimistic)~\cite{Pompa:2022cxc} &
$<0.9$ &
95\% &
10 kpc &
$<4.17\times 10^{-1}$ \\
Galactic CCSN (DUNE, conservative)~\cite{Pompa:2022cxc} &
$<2.1$ &
95\% &
10 kpc &
$<2.27$ \\
Galactic CCSN (JUNO)~\cite{Parker:2023cos} &
$<0.39$ &
95\% &
10 kpc &
$<7.82\times 10^{-2}$ \\

\hline\hline
\end{tabular}
\renewcommand{\arraystretch}{1.0}
\end{table*}

The phenomenological input to our translation consists of published or projected supernova ToF constraints, usually quoted as upper limits on an effective neutrino mass for an assumed source distance. In the present framework, these results enter through the corresponding upper limit on the leading dispersion coefficient, \(A_{\rm lim}\), obtained from Eq.~\eqref{eq:Alim_from_mlim}. The translated bound on the neutrino millicharge then follows from Eqs.~\eqref{eq:Aq_le_Alim} and \eqref{eq:Aq_total}, namely
\begin{equation}
|q_\nu|
\le
\left[\frac{A_{\rm lim}}{\mathcal{F}(L,\ell,b)}\right]^{1/2}\, ,
\label{eq:q_bound_master}
\end{equation}
where the denominator is defined by
\begin{equation}
\mathcal{F}(L,\ell,b)=
\frac{L^3}{24c}\,B_{\rm eff}^2
+
\frac{L^2\lambda_B^{\rm turb}}{9c}\,\big(B_{\rm rms}^{\rm turb}\big)^2
+
\frac{L^2\lambda_B^{\rm str}}{6c}\,\mathcal{B}_{\rm str}^2\, .
\label{eq:Fkernel_def}
\end{equation}
Here \((\ell,b)\) denote Galactic longitude and latitude, respectively. We refer to \(\mathcal{F}(L,\ell,b)\) as the \textit{magnetic delay kernel}, or simply \emph{delay kernel} hereafter, and report the final result in terms of the millicharge fraction \(\epsilon_\nu\equiv q_\nu/e\).

The remaining ingredient is therefore the set of published or projected ToF mass limits entering the translation. Supernova ToF analyses do not measure an absolute propagation time. Rather, they constrain energy dependent distortions of the observed burst time structure, encoded in the time and energy distribution of the detected neutrino signal. In practice, the unknown emission time origin is absorbed into one or more offset time nuisance parameters, so the sensitivity arises from relative timing information within the burst. This point is especially important for SN1987A, where the clocks of different detectors were not synchronized, but it also underlies modern forecasts for a future Galactic core-collapse supernovae (CCSNe).

The literature inputs used in this work span two broad categories. The first consists of \emph{model-independent} bounds, which infer a conservative limit on the leading propagation-induced dispersion from coarse burst properties such as the observed duration and energy spread. The second consists of \emph{model-dependent} analyses or sensitivity forecasts, in which the detected time and energy information are combined with an assumed emission model, detector response, and nuisance-parameter treatment. In practice, the numerical reach of the latter depends not only on the statistical framework adopted, but also on which temporal structures of the neutrino signal are effectively resolved and used in the analysis. In this context, for SN1987A we consider both the order-of-magnitude limit based on the overall signal structure~\cite{Schramm:1989rf} and the Bayesian analysis of Ref.~\cite{Loredo:2001rx}, which exploits the observed event times and energies within a specific emission model \cite{dosSantos:2021egl}. As for future prospects, the projected Super-Kamiokande sensitivity of Ref.~\cite{Pagliaroli:2010ik} is based on an improved version of the same general method, the IceCube forecast of Ref.~\cite{Ellis:2012ji} relies on millisecond-scale time variations, the DUNE forecast of Ref.~\cite{Pompa:2022cxc} is tied to the neutronization burst, and the JUNO liquid-scintillator study of Ref.~\cite{Parker:2023cos} exploits the sharp cutoff associated with black-hole formation together with the detector’s energy resolution to probe the energy-dependent ToF delay.

Table~\ref{tab:tof_inputs} summarizes the literature inputs adopted in this work by listing the quoted mass limit, the corresponding confidence level when available, the assumed source distance, and the inferred value of \(A_{\rm lim}\).

\section{Results}
\label{sec:results}

With the translation framework established in Sec.~\ref{sec:formalism}, published and projected supernova ToF constraints can now be recast into bounds on the neutrino millicharge. The results depend on two ingredients: the Galactic magnetic field kernel, which controls the propagation induced delay for a given line-of-sight, and the experimental sensitivity to the common $E_\nu^{-2}$ dispersion coefficient. We therefore begin by characterizing the magnetic kernel and its directional dependence and then present the translated limits.


\begin{figure*}[t]
\centering
\includegraphics[width=0.48\textwidth]{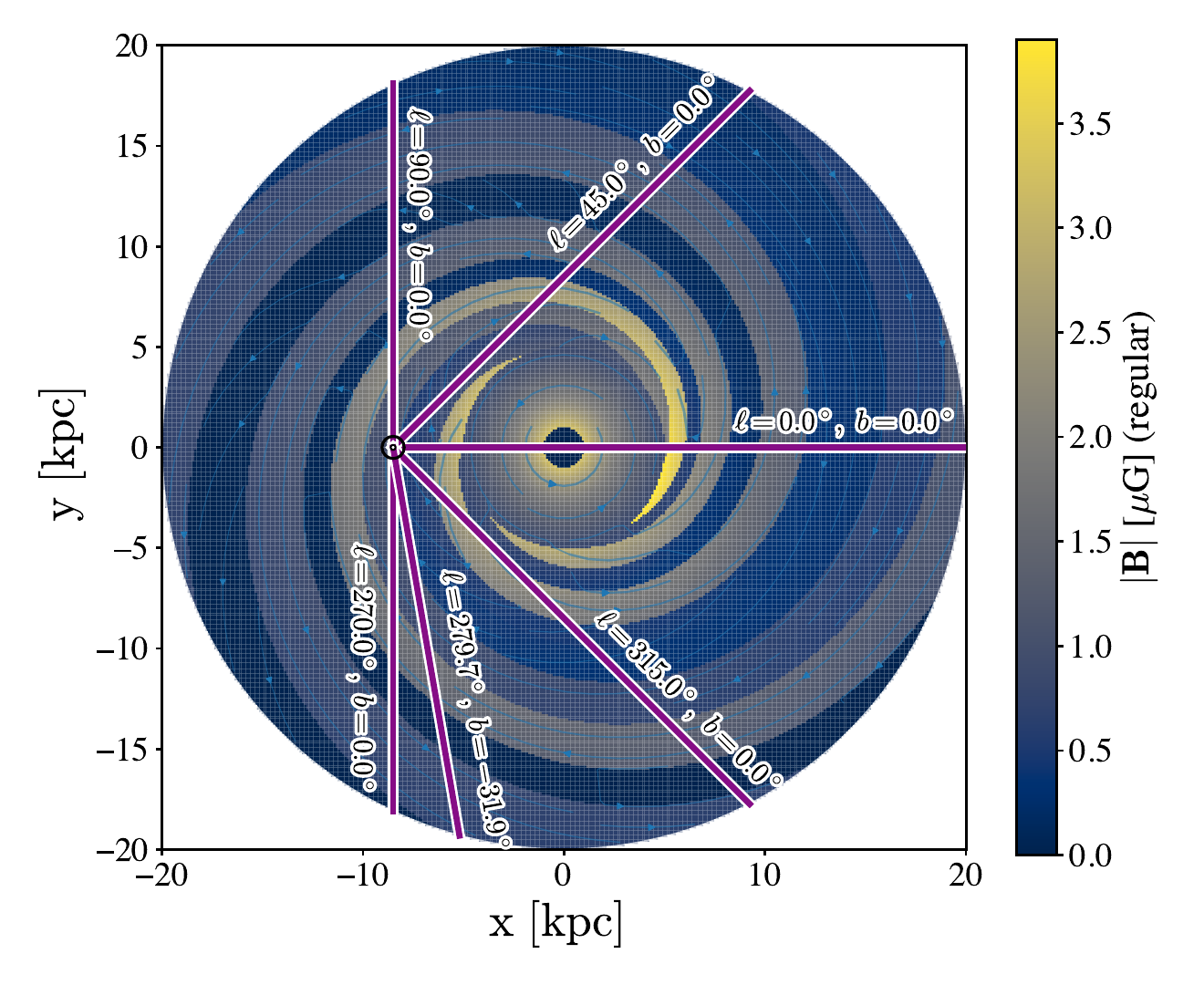}
\hfill
\includegraphics[width=0.5\textwidth]{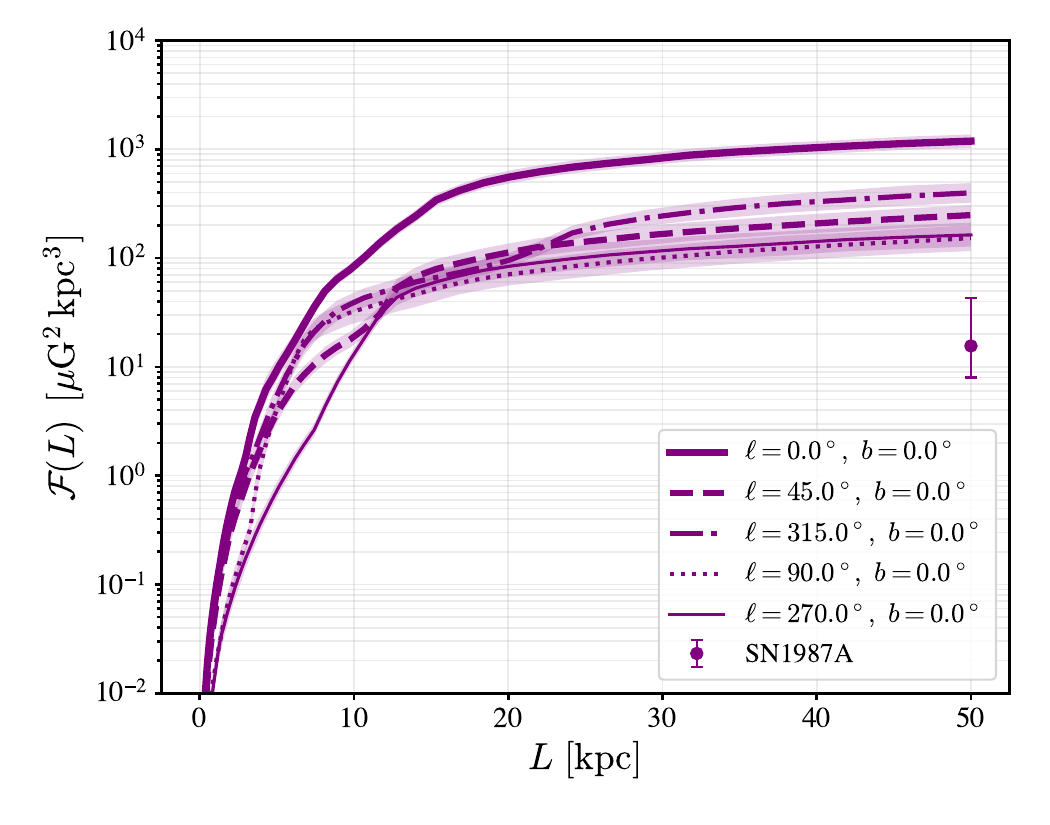}
\caption{(Left) Galactic plane map of the regular JF12 magnetic field model in Galactocentric coordinates, with the selected lines of sight used for the $\mathcal{F}$ curves shown in the right panel. The $\odot$ illustrates the Earth's position (Right) Magnetic kernel $\mathcal{F}(L,\ell,b)$ along the lines of sight indicated in the left panel. Shaded regions show the propagated $1\sigma$ band from the JF12 parameter uncertainties. The marker corresponds to the SN1987A direction ($\ell=279.7^\circ$, $b=-31.9^\circ$) evaluated at the source distance, $50$ kpc.}
\label{fig:kernel_los}
\end{figure*}


\subsection{Galactic field kernel and line-of-sight dependence}
\label{subsec:kernel_los}

The conversion from $A_{\rm lim}$ to a bound on $\epsilon_\nu$ is controlled by the effective magnetic kernel $\mathcal{F}(L,\ell,b)$, defined in Eq.~\eqref{eq:Fkernel_def}. This kernel encodes the cumulative impact of the GMF on the propagation of a millicharged neutrino from the supernova source to the Solar position. The translated limits therefore depend not only on the overall field strength, but also on the direction dependent path through the GMF and on the relative importance of the regular and stochastic contributions entering the kernel.

Figure~\ref{fig:kernel_los} illustrates this dependence in two complementary ways. The left panel displays the regular JF12 GMF configuration in the Galactic plane, together with representative lines-of-sight. The right panel shows the corresponding behavior of the $\mathcal{F}(L)$ as a function of path length, including the propagated uncertainty associated with the JF12 parameters. These curves make clear that the translation from ToF sensitivity to millicharge reach depends strongly on direction: some sightlines accumulate a substantially larger delay kernel and therefore yield stronger bounds on $\epsilon_\nu$, whereas others are comparatively weak and lead to less restrictive constraints. The SN1987A direction is also shown for direct comparison.

In practice, the delay kernel is typically enhanced for sightlines that remain close to the Galactic disk, since such trajectories sample the regular large scale GMF over a longer fraction of the propagation path. Even within the plane, however, the kernel is not uniform: the examples shown in Figure~\ref{fig:kernel_los} indicate that changing the Galactic longitude can modify $\mathcal{F}(L,\ell,b)$ by a factor of a few, while the direction toward the Galactic center ($\ell=0^\circ$, $b=0^\circ$) can yield a substantially larger kernel, approaching an order of magnitude enhancement relative to less favorable in-plane directions at large distances.

A caveat concerns directions approaching the Galactic center. In the original JF12 implementation, the regular field is manually set to zero in the innermost region ($r<1$ kpc), so the model is not strictly divergence free there. A solenoidal improvement of JF12 has been developed in Ref.~\cite{1987PhRvD..36.3276L}, where the regular disk and X-field components are modified to satisfy $\nabla\!\cdot\!\mathbf{B}=0$ more consistently. In addition, dedicated magnetic field models exist for the central molecular zone (CMZ), i.e.\ the inner $\lesssim 200$ pc of the Galaxy, where the field structure is known to differ from the large scale disk and halo components \cite{Guenduez:2019cwe}. In this work, however, we adopt the original JF12 model throughout in order to retain a single, conservative baseline GMF model throughout the analysis. In our context, these considerations mainly affect the interpretation of directions very close to $(\ell,b)=(0^\circ,0^\circ)$. For such lines of sight and for the distance $L=10$ kpc relevant to our translated Galactic CCSN benchmarks, using a solenoidal JF12 realization enhances the field kernel by $\sim 60\,\%$, corresponding to a strengthening of the translated millicharge bound by $\sim 20 \,\%$. Considering a CMZ scale component is expected to have a smaller impact on the quantities of interest here, since the spatial extent of the central region is small compared with the Galactic propagation length entering the kernel.

\begin{figure}[t]
  \centering
  \includegraphics[width=0.99\columnwidth]{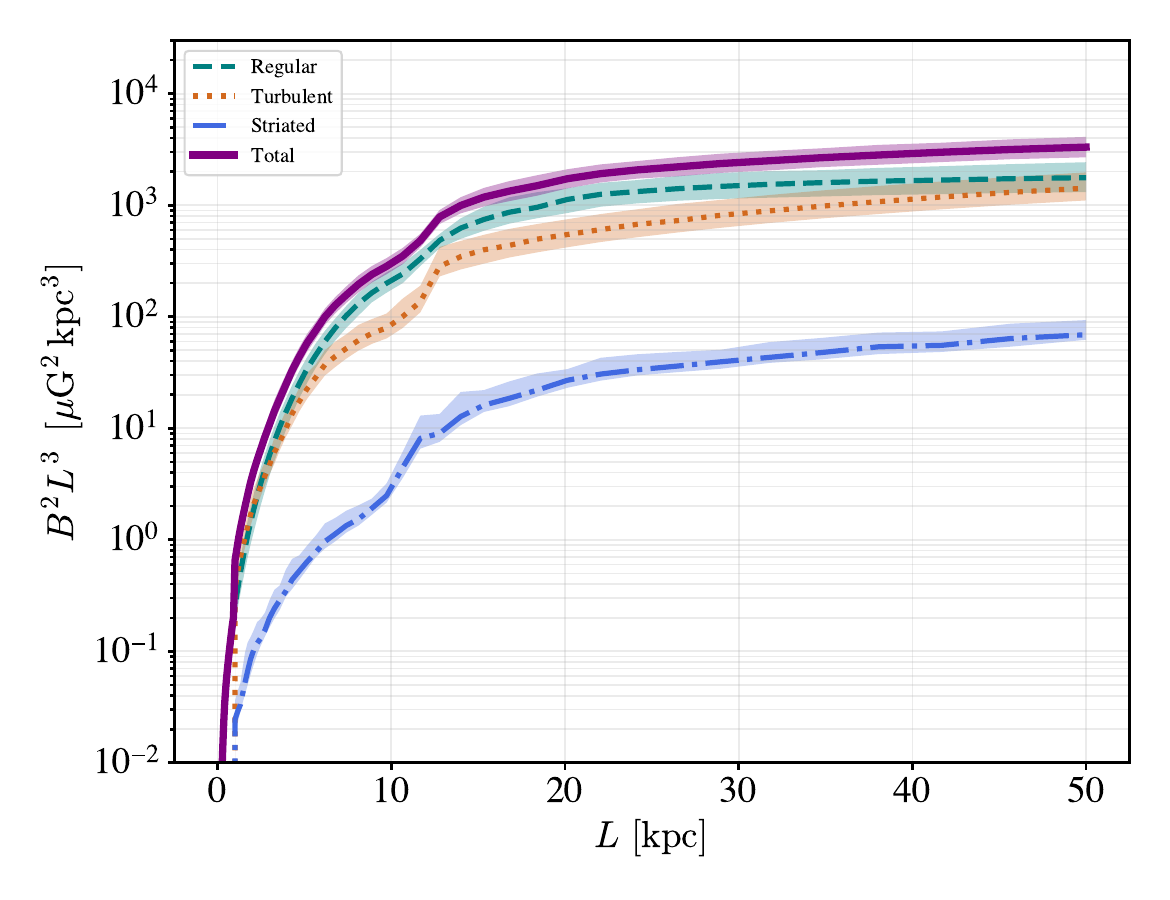}
  \caption{
  Decomposition of the magnetic kernel building blocks for the benchmark direction $(\ell,b)=(45^\circ,0^\circ)$ as a function of source distance. We show the regular contribution through $B_{\rm eff}^2L^3$ and the stochastic contributions through $B_{\rm rms}^2\lambda_B L^2$ for the turbulent and striated components. The striated term is consistently subdominant, while the regular and turbulent terms are both relevant for the total kernel. Although the regular quantity is systematically larger, the different prefactors entering Eq.~\eqref{eq:Fkernel_def} make the regular and turbulent contributions to the final kernel comparable in magnitude.
  }
  \label{fig:field_decompose_l45b0}
\end{figure}

A complementary view is provided by the decomposition shown in Figure~\ref{fig:field_decompose_l45b0} for the benchmark direction $(\ell,b)=(45^\circ,0^\circ)$. This example also reflects the broader trend found in our scan: the striated contribution is consistently subdominant, remaining about one to two orders of magnitude below the regular and turbulent terms over the full distance range considered. By contrast, the regular and turbulent components are both important for the total kernel: although the regular quantity $B_{\rm eff}^2L^3$ is systematically larger than the turbulent combination $B_{\rm rms}^2\lambda_B L^2$, their different prefactors in the delay kernel, $1/24$ and $1/9$, make their final contributions to $\mathcal{F}$ comparable in magnitude.

The SN1987A sightline is qualitatively different. Because it lies significantly off the Galactic plane ($b=-31.9^\circ$), the neutrino trajectory samples a smaller portion of the strong large scale GMF before leaving the disk, and the remaining propagation occurs predominantly outside the Milky Way, where the magnetic field is expected to be much weaker. As a result, the associated delay kernel is suppressed by more than an order of magnitude relative to directions close to the Galactic plane, as seen in the right panel of Figure~\ref{fig:kernel_los}. This suppression is accompanied by a comparatively larger relative uncertainty band. The reason is that, for such an off-plane trajectory, the kernel receives a larger fraction of its contribution from the toroidal halo and poloidal X-field components, rather than from the disk field. The disk field is more directly anchored by the stronger rotation measure and synchrotron signal near the Galactic plane, while the out of plane components are comparatively more sensitive to the global modeling assumptions entering the fit. As a result, the uncertainty on $\mathcal{F}$ is larger for off-plane directions than for typical in-plane line-of-sights.


\begin{table*}[t]
\centering
\caption{
Translated limits on the neutrino millicharge obtained by combining the ToF inputs of Table~\ref{tab:tof_inputs} with the Galactic magnetic field kernel in the JF12 framework. We report the benchmark bound for the reference in-plane direction $(\ell,b)=(45^\circ,0^\circ)$ together with its propagated $1\sigma$ GMF uncertainty, as well as the directional spread of the central prediction over the scanned grid $-5^\circ<b<5^\circ$, $0^\circ<\ell<360^\circ$, summarized by the percentile interval $[\epsilon_{\nu,\rm p5},\epsilon_{\nu,\rm p95}]$. For completeness, we also quote the strongest translated bound found in the scanned grid; this value should be interpreted as an optimistic directional extreme rather than a benchmark result.
}
\label{tab:translated_limits}
\renewcommand{\arraystretch}{1.35}
\begin{tabular}{lcccc}
\hline\hline
\textbf{Analysis (Ref.)} &
\begin{tabular}[c]{@{}c@{}}\textbf{Benchmark}\\ \textbf{$|\epsilon_\nu|/10^{-20}$}\end{tabular} &
\begin{tabular}[c]{@{}c@{}}\textbf{$1\sigma$}\\ \textbf{GMF}\end{tabular} &
\begin{tabular}[c]{@{}c@{}}\textbf{Directional spread}\\ \textbf{$[\epsilon_{\nu,\rm p5},\epsilon_{\nu,\rm p95}] /10^{-20}$}\end{tabular} &
\begin{tabular}[c]{@{}c@{}}\textbf{Strongest grid bound}\\ \textbf{$|\epsilon_\nu|/10^{-20}$}\end{tabular} \\
\hline
SN1987A (model-indep.)~\cite{Schramm:1989rf} &
$4138.5$ &
${}^{+1663}_{-1657}$ &
--- &
--- \\
SN1987A (model-dep.)~\cite{Loredo:2001rx} &
$785.6$ &
${}^{+315.9}_{-314.5}$ &
--- &
--- \\
\hline
Galactic CCSN (SK)~\cite{Pagliaroli:2010ik} &
$44.5$ &
${}^{+4.0}_{-3.7}$ &
$[20.2,\,67.3]$ &
17.1 \\
Galactic CCSN (IceCube)~\cite{Ellis:2012ji} &
$7.8$ &
${}^{+0.7}_{-0.6}$ &
$[3.54,\,11.8]$ &
3.0 \\
Galactic CCSN (DUNE, optimistic)~\cite{Pompa:2022cxc} &
$50.1$ &
${}^{+4.5}_{-4.1}$ &
$[22.8,\,75.8]$ &
19.3 \\
Galactic CCSN (DUNE, conservative)~\cite{Pompa:2022cxc} &
$116.8$ &
${}^{+10.5}_{-9.7}$ &
$[53.2,\,176.9]$ &
44.9 \\
Galactic CCSN (JUNO)~\cite{Parker:2023cos} &
$21.6$ &
${}^{+1.9}_{-1.8}$ &
$[9.8,\,32.8]$ &
8.3 \\

\hline\hline
\end{tabular}
\renewcommand{\arraystretch}{1.0}
\end{table*}


\subsection{Limits on the neutrino millicharge}
\label{subsec:translated_limits}

We summarize the translated limits in two complementary ways. First, we quote a benchmark result for the in-plane direction $(\ell,b)=(45^\circ,0^\circ)$, together with the associated $1\sigma$ uncertainty obtained by propagating the uncertainty on the full delay kernel $\mathcal{F}$. We adopt this line-of-sight because it is favorable yet non-extremal, and it avoids the additional modeling ambiguity associated with the innermost Galactic center region. Second, we quantify the directional spread of the central prediction by scanning a grid over $-5^\circ<b<5^\circ$ and $0^\circ<\ell<360^\circ$ at fixed $L=10~{\rm kpc}$. This restricted scan is intended as an illustration of how the GMF translation varies across near-plane line-of-sights relevant to Galactic CCSNe. Extending the scan to the full sky would mainly broaden the quoted directional interval by introducing additional off-plane sightlines with smaller magnetic kernels and hence weaker translated bounds, while leaving the favorable near-plane end essentially unchanged. From this restricted sky scan we extract the percentile bounds $\epsilon_{\nu,\rm p5}$ and $\epsilon_{\nu,\rm p95}$, which enclose 90\% of the sampled directions. 

Thus, we probe two different aspects of the problem: the benchmark result characterizes the propagated field model uncertainty for one representative in-plane sightline, while the percentile interval captures the directional variation of the central prediction across the near-plane region.

For completeness, we also identify the strongest translated bound found within our scanned grid. This value should be interpreted with caution, however, since it depends on the adopted GMF model and is influenced by directions approaching the poorly constrained Galactic center region. We therefore use it only as an optimistic directional extreme, rather than as the benchmark result. Table~\ref{tab:translated_limits} summarizes the translated millicharge limits corresponding to the ToF inputs of Table~\ref{tab:tof_inputs}.

A few remarks are in order. First, for fixed source distance, the spread of translated bounds across the scanned grid is controlled entirely by the distribution of the delay kernel $\mathcal{F}(L,\ell,b)$, while the ToF input fixes only the overall normalization. Since $|\epsilon_\nu|_{\rm lim}\propto \mathcal{F}^{-1/2}$, the directional spread of the translated limits corresponds to the square root of the inverse kernel spread; in our scanned grid, this amounts to a factor of about $\sim 3$ in $|\epsilon_\nu|_{\rm lim}$. Numerically, for the projected Galactic burst sensitivities considered here, this astrophysical spread is comparable to the differences among nearby detector forecasts. Second, the translated SN1987A bounds remain substantially weaker than the Galactic forecasts. This reflects both the limited statistics of the SN1987A signal and the fact that its off-plane direction samples a much smaller delay kernel than a typical near-plane CCSN.

For DUNE, the literature quotes a range of projected mass sensitivities rather than a unique value. We therefore retain both optimistic and conservative entries corresponding to the endpoints of the quoted range, rather than compressing them into a single representative number. For future Galactic CCSNe observations, the most favorable translated bounds arise when strong ToF sensitivity is combined with a favorable in-plane line-of-sight; in such cases, the translated reach can enter the low-$10^{-19}\, e$ regime and, for the most favorable directions in our scan, approach the $10^{-20}\,e$ level.

\subsection{Variation with the GMF models}
The limits derived in this section depend on the adopted GMF model. To quantify this dependence, we repeat the calculation of the regular-field contribution using the recent Unger--Farrar regular GMF models, hereafter UF23~\cite{Unger:2023lob}. As in JF12, the UF23 models constrain the coherent GMF using the Faraday rotation measures of extragalactic radio sources and the polarized synchrotron intensity of Galactic cosmic-ray electrons, however, employing new divergence-free parametrizations of the large-scale field and updated full-sky rotation measure and polarized-intensity maps. The suite also explores different assumptions about the thermal-electron and cosmic ray electron distributions entering the rotation measure and synchrotron predictions. Based on these assumptions eight benchmark models, \texttt{base}, \texttt{neCL}, \texttt{expX}, \texttt{spur}, \texttt{cre10}, \texttt{synCG}, \texttt{twistX}, and \texttt{nebCor}, were selected to span the present modeling uncertainties in the regular GMF. 

\begin{figure}[t]
  \centering
  \includegraphics[width=0.97\columnwidth]{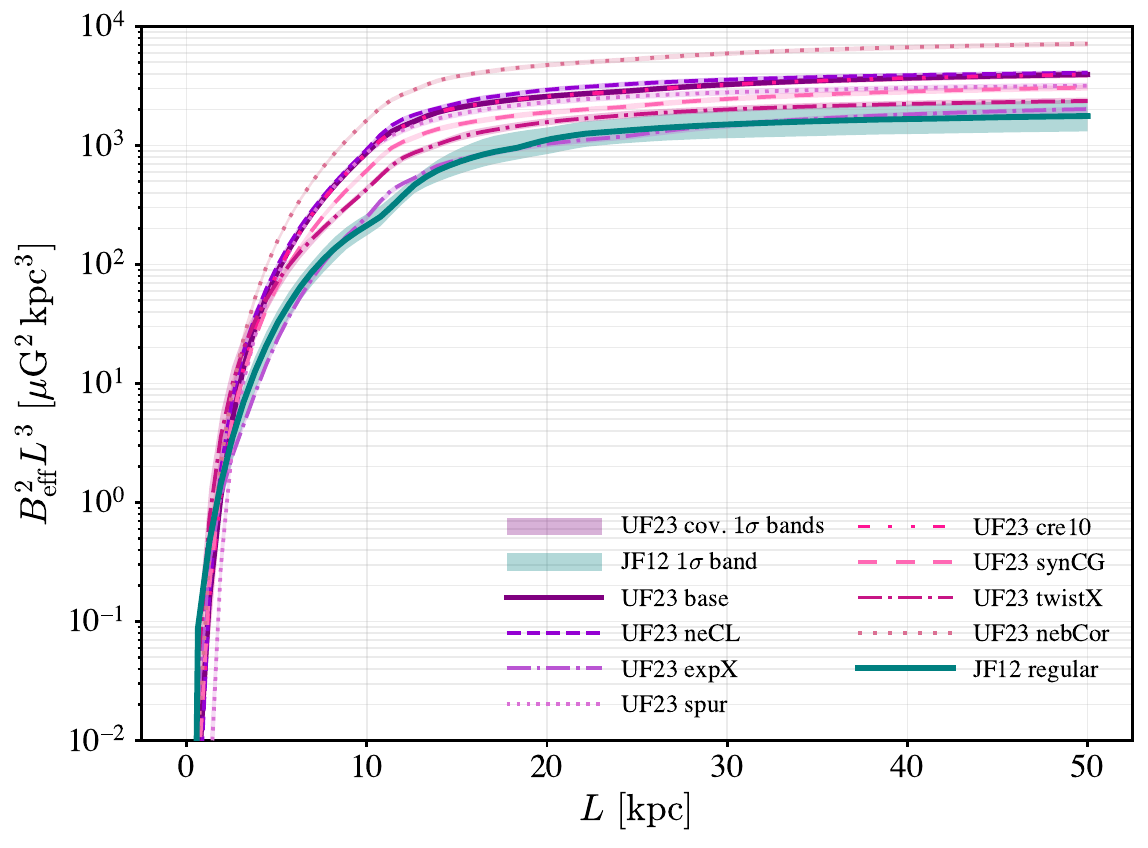}
  \label{fig:uf23_benchmark}
  \includegraphics[width=0.99\columnwidth]{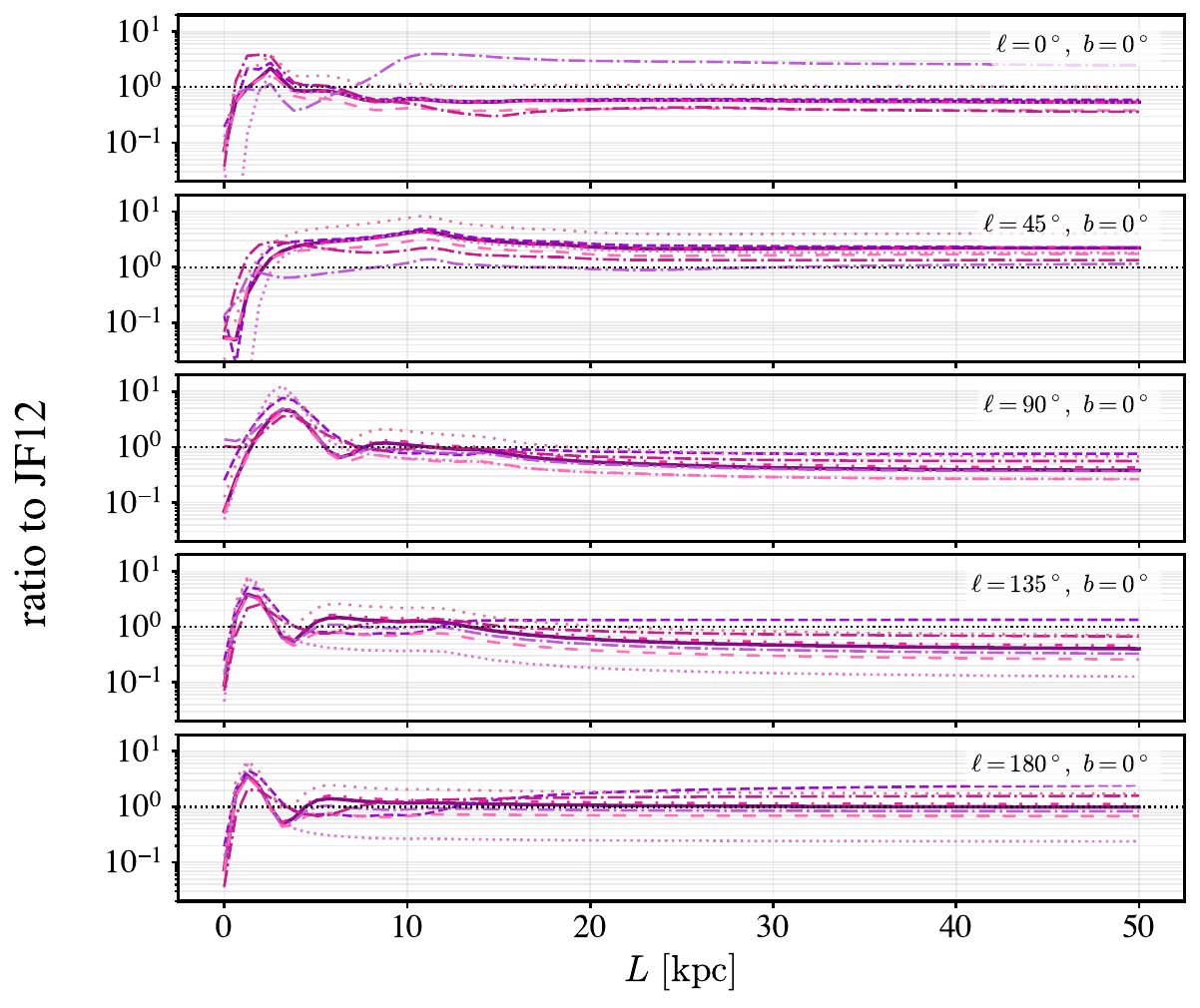}
  \label{fig:uf23_jf12_ratio}
  \caption{(Top) The quantity \(B_{\rm eff}^2L^3\) for the benchmark direction $(\ell,b)=(45^\circ,0^\circ)$. The teal curve shows the JF12 reference baseline, together with its phenomenological propagated  $1\sigma$ band. The purple/pink curves show the eight UF23 benchmark coherent-field models. The UF23 shaded regions show the propagated \(1\sigma\) bands obtained from the model-specific covariance matrices supplied with the public UF23 implementation. In all cases, the central curve corresponds to the best-fit prediction. (Bottom) Ratio of \(B_{\rm eff}^2L^3\) obtained with the eight UF23 benchmark models to the corresponding JF12 result. The panels show directions in the Galactic plane as in Figure \ref{fig:kernel_los}. The same colors and line styles as in top panel are used for the UF23 models.}
  \label{fig:allmodels_l45b0}
\end{figure}

For each UF23 benchmark model, we compute the same quantity used in the baseline analysis, $B_{\rm eff}^2L^3$, along the line-of-sight. The result is shown in Figure \ref{fig:allmodels_l45b0} for the benchmark direction $(\ell,b)=(45^\circ,0^\circ)$, together with the corresponding JF12 regular component and its phenomenological propagated $1\sigma$ band described in
Appendix~\ref{app:gmf}. For the UF23 models, the shaded bands are obtained by propagating the model-specific covariance matrices supplied with the public
UF23 implementation through the full \(B_{\rm eff}^2L^3\) calculation.\footnote{ \url{https://zenodo.org/records/11321212}, also see Appendix C in \cite{Unger:2023lob}} The lines show the best-fit prediction of each model.

Several features are apparent from Figure \ref{fig:allmodels_l45b0}. First, the
spread among the UF23 benchmark models is larger than the within-model covariance band for a fixed UF23 model. This indicates that, for the present observable, the dominant coherent-field uncertainty is associated with the choice of global GMF reconstruction rather than with the local covariance of the parameters within a given reconstruction. Second, the UF23/JF12 ratio depends on both distance and direction. For example, for the benchmark direction
$(\ell,b)=(45^\circ,0^\circ)$ and $L=10\,{\rm kpc}$, some UF23 models give $\frac{(B_{\rm eff}^2L^3)_{\rm UF23}}{(B_{\rm eff}^2L^3)_{\rm JF12}}\simeq 8$. This corresponds to a stronger limit by a factor $\lesssim 2.8$ on the $\epsilon_\nu$. 
More generally, the regular-field contribution to $B_{\rm eff}^2L^3$, and therefore to the magnetic delay kernel, can be either
enhanced or reduced, by a factor of $\sim$ few, relative to the JF12 reference, depending on the assumed coherent GMF model and on the Galactic line-of-sight. These variations lead to more stringent or weaker bounds, respectively. The directional comparison in the bottom panel of Figure \ref{fig:allmodels_l45b0} illustrates this model dependence more directly.

We keep JF12 as the baseline GMF model for the full analysis because it provides the regular, striated, and turbulent components used in our treatment, whereas UF23 updates only the regular field and its parameter covariance. The UF23 comparison therefore quantifies the model dependence of the regular field contribution alone.

\section{Discussion and Conclusion}
\label{sec:discussion}

The translated bounds derived above may be viewed in the broader landscape of existing constraints on neutrino millicharge.

Direct bounds on neutrino millicharge come from scattering measurements, where a nonzero charge would modify the low-recoil event spectrum. For reactor neutrinos, the strongest limits reach the \(\sim 10^{-12}\,e\) level \cite{Gninenko:2006fi, Studenikin:2013my, Chen:2014dsa}. Accelerator-based experiments probe other flavors, with CE\(\nu\)NS/COHERENT analyses constraining the muon-neutrino millicharge at the \(\sim 10^{-10}\,e\) level \cite{AtzoriCorona:2022qrf, Das:2020egb}. More recently, solar-neutrino scattering in xenon-based dark-matter detectors has pushed this class substantially further, yielding constraints at the \(\sim 10^{-13}\,e\) level \cite{A:2022acy}, depending on the flavor assumptions and treatment of atomic effects.

A second class of bounds comes from stellar-cooling arguments. There, the relevant effect is the enhancement of photon-mediated plasma processes, most notably plasmon decay to $\nu\bar{\nu}$, which would increase the stellar energy-loss rate. Such considerations typically lead to limits at the \(\sim 10^{-14}\,e\) level, with some dependence on the stellar environment and adopted cooling criterion \cite{Bernstein:1963qh, Raffelt:1999gv, Giunti:2014ixa}. 

The strongest commonly quoted benchmark is instead an indirect constraint, obtained from laboratory tests of the electrical neutrality of matter together with bounds on the neutron charge and the assumption of charge conservation in \(\beta\) decay. This implies a neutrino-charge constraint at the level of \(|q_\nu|\lesssim {\rm few}\times 10^{-21}\,e\)~\cite{Raffelt:1999gv, Giunti:2014ixa}. \footnote{A recent reassessment has argued that the much stronger \(10^{-35}\,e\)-level bound sometimes quoted in the literature is based on a controversial cosmological argument and should not be regarded as comparably robust. In that sense, the laboratory neutrality-based bound remains the relevant benchmark for comparison with direct or phenomenological probes~\cite{Karshenboim:2024iff}}.

Moreover, an astrophysical bound is provided by the ``neutrino star turning'' mechanism in rotating magnetized neutron-star matter, which yields a constraint of order \(|q_\nu|\lesssim 10^{-19}\,e\) from the requirement that escaping millicharged neutrinos not induce an excessive change in the observed pulsar rotation rate~\cite{Studenikin:2012vi}. That result is derived in a specific dense matter setting namely neutrino propagation in rotating magnetized neutron-star matter, where the relevant observable is the torque exerted on the stellar medium by the escaping neutrino flux.

\begin{figure}
    \centering
    \includegraphics[width=\linewidth]{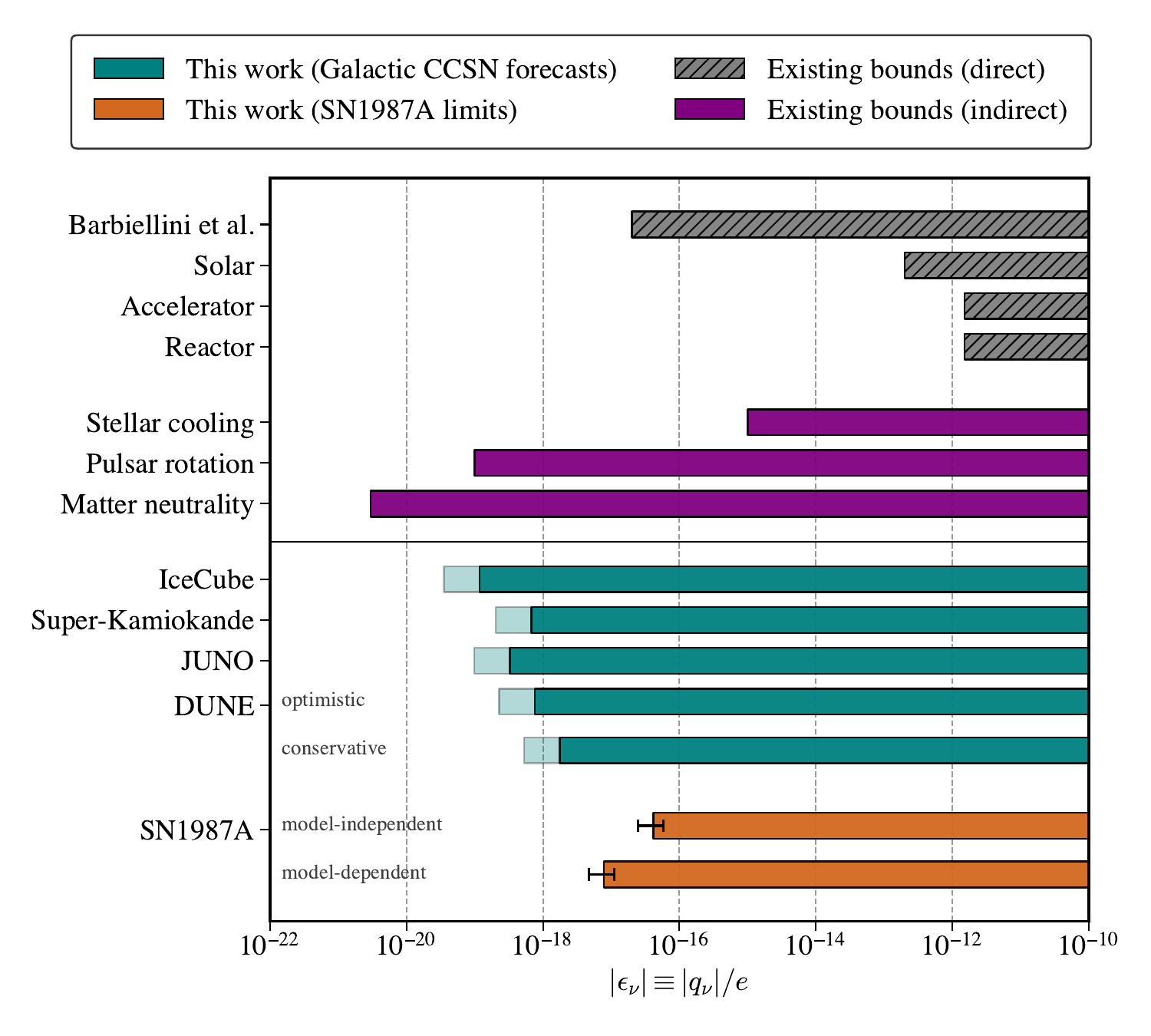}
    \caption{Comparison of neutrino-millicharge bounds from different probes, with the excluded region lying to the right. Lower panel shows the limits derived in this work: SN1987A constraints (orange) and projected sensitivities for a future Galactic CCSNe (green). For the Galactic forecasts, the lighter extension at the end of each bar indicates the variation across the percentile interval reported in Table~\ref{tab:translated_limits}; for SN1987A, the error bars represent the propagated GMF systematic uncertainty. Upper panel shows the existing bounds in the literature. Hatched gray bars denote direct probes, including the earlier SN1987A ToF estimate of ~\cite{Barbiellini:1987zz} and scattering constraints from reactor, accelerator, and solar neutrino measurements \cite{Gninenko:2006fi, Studenikin:2013my, AtzoriCorona:2022jeb, A:2022acy}. Purple bars denote indirect or model-dependent bounds, including the neutrality of matter \cite{Raffelt:1999gv}, the pulsar rotation constraint \cite{Studenikin:2012vi}, and stellar cooling limits \cite{Raffelt:1999gv, Fung:2023euv}.}
    \label{fig:q_nu_limits_comparison}
\end{figure}

In this context, the supernova limits derived in this work probe neutrino millicharge through a qualitatively distinct observable: energy dependent distortions in the time profile of a transient neutrino signal after astrophysical distance propagation through the adopted GMF model. In this sense, they are complementary to laboratory scattering, stellar-cooling, neutrality-based, and more specialized astrophysical arguments. Unlike the pulsar-rotation bound discussed above, the present translation is formulated directly in terms of the effective millicharge governing propagation through the GMF, and is therefore less tied to the detailed microphysics of a particular dense matter environment. Figure~\ref{fig:q_nu_limits_comparison} summarizes these bounds together with the translated SN1987A limits and the projected Galactic-core-collapse-supernova sensitivities derived in this work.

A few additional remarks help clarify how the translated bounds should be interpreted.

(\textit{i}) the Galactic forecasts quoted in Table~\ref{tab:tof_inputs} are benchmark sensitivities evaluated at \(L=10~{\rm kpc}\), while the SN1987A entries correspond to \(L\simeq 50~{\rm kpc}\) as adopted in the original analyses. For a source at a different distance, the translated charge bound changes through two distinct effects: the delay kernel \(\mathcal{F}(L,\ell,b)\) changes with the propagation length and direction, while the underlying ToF sensitivity \(A_{\rm lim}\) changes because both the event statistics and the resolvability of the relevant temporal structures depend on distance in a detector- and analysis-specific way. The former effect can be recomputed directly within our framework and, as illustrated in Fig.~\ref{fig:kernel_los}, generally enhances the propagation kernel with increasing distance. For the representative in-plane sightlines, we find \(\mathcal{F}(25~{\rm kpc})/\mathcal{F}(10~{\rm kpc})\sim 2\)–\(9\), which at fixed \(A_{\rm lim}\) would strengthen the translated millicharge bound by a factor \(\sqrt{\mathcal{F}(25)/\mathcal{F}(10)}\sim 1.4\)–\(3\). The detector-side dependence, however, is not captured by a universal distance rescaling. As one explicit example, the DUNE analysis of Ref.~\cite{Pompa:2022cxc} finds that the mass sensitivity worsens by about a factor of $\sim 2$ as the source distance increases from \(5\) to \(25~{\rm kpc}\). Combining this with the propagation-side enhancement above suggests that, for such a case, the translated charge reach may improve only modestly with distance, rather than following a simple monotonic scaling. More generally, precise distance-dependent millicharge limits require the corresponding distance-dependent ToF analysis for each experiment.

(\textit{ii}) In the parameter region relevant here, the net magnetic deflection remains extremely small, so the small-angle approximation underlying the magnetic-delay derivation is well satisfied. In this regime, the delay is dominated by the geometric propagation length increase, while the associated angular shift is far too small to be resolved by MeV supernova neutrino detectors.

(\textit{iii}) The translation developed in this work is most straightforward in the null-delay regime, where the observed ToF data are interpreted as setting an upper bound on a common leading \(E_\nu^{-2}\) dispersion coefficient. In that case, the standard mass-induced delay and the millicharge-induced magnetic delay are degenerate at the level of their leading energy dependence, and the allowed timing budget may be reinterpreted as an effective millicharge contribution. If the standard mass-induced delay is itself non-negligible at the relevant level of sensitivity, however, the translation to \(q_\nu\) is no longer independent and should instead be treated in a joint analysis.

(\textit{iv}) A further subtlety is that a supernova detector does not, in general, observe a single mass eigenstate, but rather a signal built from an incoherent mixture of neutrino mass eigenstates. Accordingly, a ToF analysis admits a direct reinterpretation in terms of a single millicharge bound only when it constrains a single effective leading $E_\nu^{-2}$ dispersion coefficient for the detected signal. By contrast, once the analysis becomes sensitive to the individual neutrino mass eigenstates, there is no longer, in general, a unique single coefficient to translate. In that regime, the mapping to neutrino millicharge becomes model dependent, because one must specify how the electric charge interaction is realized in the mass basis. A direct eignenstate-by-eigenstate translation is straightforward only under additional assumptions, such as a diagonal charge assignment for the propagating mass eigenstates, while a more general treatment would require a dedicated analysis and is beyond the scope of this work. This issue is relevant, for example, to the JUNO forecast of Ref. \cite{Denton:2024mlb}, which is formulated in terms of sensitivity to the individual neutrino mass eigenstates rather than to a single effective dispersion coefficient; accordingly, we do not include such results among the direct inputs to our translated bounds. 

Within this scope, the translated millicharge bounds inherit the assumptions of the underlying ToF analyses, including the adopted GMF model, the adopted supernova emission model, detector thresholds and resolutions, nuisance-parameter treatment, and the temporal structures that remain experimentally accessible.
The analysis of Ref. \cite{Barbiellini:1987zz} established the essential time-of-flight idea underlying this class of probes. The present framework recasts that idea in the common \(A_{\rm ToF}\) language and replaces the original uniform field estimate with a line-of-sight dependent translation based on the adopted GMF model. In this sense, SN1987A serves both as a historical benchmark and as a genuine data based constraint, while the strongest prospective reach of the method is provided by a future high statistics Galactic CCSNe. Among existing bounds, the most demanding benchmark remains the neutrality of matter limit: surpassing it would require sensitivity an exceptionally small allowed \(E_\nu^{-2}\) dispersion coefficients, and hence to very fine temporal structure in the supernova signal, whose physical presence and experimental resolvability both remain uncertain.

Subject to these qualifications, the framework developed here provides a map from supernova ToF sensitivity to neutrino millicharge sensitivity, complementary to the existing laboratory, stellar, and indirect bounds.

\begin{acknowledgments}

We thank Orlando L.G. Peres, Arman Esmaili, and Renata Zukanovich Funchal for their valuable comments on the manuscript.
P. D. N. acknowledges support from the São Paulo Research Foundation (FAPESP) through a postdoctoral fellowship, Grant No. 2025/13010-1. A.F. E. acknowledges support from the São Paulo Research Foundation (FAPESP) through a postdoctoral fellowship, Grant No. 2025/03199-0. G.A. N. acknowledges support from the São Paulo Research Foundation (FAPESP) through a Ph.D. scholarship No. 2024/01465-1. 

\end{acknowledgments}

\appendix

\section{Derivation of the magnetic-delay kernel}
\label{app:kernel_derivation}

In this appendix, we derive the line-of-sight expression used in Eq.~\eqref{eq:dtq_reg_kernel} for the small-angle time delay of a millicharged neutrino propagating through a prescribed transverse magnetic field. The derivation is carried out in the small-angle deflection limit.

Consider a neutrino propagating from the source at $x=0$ to the observer at $x=L$. In the absence of magnetic deflection, the trajectory is a straight line along the $x$ axis. In the small-angle regime, the actual trajectory may be written as
\begin{equation}
\mathbf{r}(x)=x\,\hat{\mathbf{x}}+\mathbf{y}(x)\, ,
\end{equation}
where $\mathbf{y}(x)$ is the two-dimensional transverse displacement, with
\begin{equation}
|\mathbf{y}'(x)| \ll 1\, .
\end{equation}
The source and observer positions are held fixed, so the transverse displacement satisfies Dirichlet boundary conditions,
\begin{equation}
\mathbf{y}(0)=\mathbf{y}(L)=0\, .
\label{eq:dirichlet_bc}
\end{equation}

The physical path length is
\begin{equation}
s=\int_0^L dx\,\sqrt{1+|\mathbf{y}'(x)|^2}\, .
\end{equation}
Expanding to leading nontrivial order in the small-angle approximation gives
\begin{equation}
s \simeq L+\frac12\int_0^L dx\,|\mathbf{y}'(x)|^2\, .
\end{equation}
Therefore the geometric arrival time delay relative to the undeflected straight path is
\begin{equation}
\Delta t \simeq \frac{s-L}{c}
= \frac{1}{2c}\int_0^L dx\,|\mathbf{y}'(x)|^2\, .
\label{eq:dt_from_yprime}
\end{equation}

Equation~\eqref{eq:dt_from_yprime} is the standard small-angle expression: the delay is controlled by the integrated squared transverse slope of the trajectory.

For an ultra-relativistic particle of charge $q_\nu$ and energy $E$, the Lorentz force induces a transverse acceleration governed by
\begin{equation}
\frac{d\hat{\mathbf{n}}}{ds} \simeq \frac{q_\nu c}{E}\,\hat{\mathbf{n}}\times \mathbf{B}\, ,
\end{equation}
where $\hat{\mathbf{n}}$ is the local propagation direction. In the small angle limit, $\hat{\mathbf{n}}\simeq \hat{\mathbf{x}}$, $ds\simeq dx$, so only the magnetic field component transverse to the unperturbed line-of-sight contributes. The transverse equation of motion may therefore be written as
\begin{equation}
\mathbf{y}''(x)=\frac{q_\nu}{E}\,\hat{\mathbf{x}}\times \mathbf{B}_\perp(x)\, .
\label{eq:ypp_force}
\end{equation}

Since $\hat{\mathbf{x}}\times$ act as a fixed orthogonal rotation in the transverse plane, it preserves the norms and inner products and therefore does not affect the final quadratic delay formula.

To solve Eq.~\eqref{eq:ypp_force} with the boundary conditions in Eq.~\eqref{eq:dirichlet_bc}, we introduce the Green's function $G(x,x')$ of the one-dimensional operator $-d^2/dx^2$ on the interval $[0,L]$,
\begin{equation}
-\frac{d^2}{dx^2}G(x,x')=\delta(x-x')\, ,
\qquad
G(0,x')=G(L,x')=0\, .
\label{eq:green_eq}
\end{equation}
The solution is
\begin{equation}
G(x,x')=
\frac{1}{L}
\begin{cases}
x(L-x')\,, & x<x'\,,\\[4pt]
x'(L-x)\,, & x>x'\,.
\end{cases}
\label{eq:green_piecewise}
\end{equation}
Equivalently,
\begin{equation}
G(x,x')=\min(x,x')-\frac{xx'}{L}\, .
\label{eq:green_min}
\end{equation}
We then have
\begin{equation}
\mathbf{y}(x)=\frac{q_\nu}{E}\int_0^L dx'\,G(x,x')\,\mathbf{B}_\perp(x')\, .
\label{eq:y_from_green}
\end{equation}

Differentiating with respect to $x$ gives
\begin{equation}
\mathbf{y}'(x)=\frac{q_\nu}{E}\int_0^L dx'\,\partial_x G(x,x')\,\mathbf{B}_\perp(x')\, .
\label{eq:yprime_from_green}
\end{equation}

Substituting Eq.~\eqref{eq:yprime_from_green} into Eq.~\eqref{eq:dt_from_yprime}, we obtain
\begin{multline}
\Delta t
=\\ 
\frac{1}{2c}\left(\frac{q_\nu}{E}\right)^2
\int_0^L dx
\int_0^L dx'\int_0^L dx'' \,
\partial_x G(x,x')\,\partial_x G(x,x'')\,\\ \times
\mathbf{B}_\perp(x')\cdot \mathbf{B}_\perp(x'')\, .
\label{eq:dt_triple}
\end{multline}
Using the identity
\begin{equation}
\int_0^L dx\, \partial_x G(x,x')\,\partial_x G(x,x'')
=
G(x',x'')\, ,
\label{eq:green_identity}
\end{equation}
which follows by integrating by parts and using Eq.~\eqref{eq:green_eq} together with the Dirichlet boundary conditions, Eq.~\eqref{eq:dt_triple} reduces to
\begin{equation}
\Delta t
=
\frac{1}{2c}\left(\frac{q_\nu}{E}\right)^2
\int_0^L dx'\int_0^L dx''\,
\mathbf{B}_\perp(x')\cdot \mathbf{B}_\perp(x'')\,
G(x',x'')\, .
\end{equation}
Renaming the dummy variables and defining
\begin{equation}
K(x,x')\equiv G(x,x')
=
\min(x,x')-\frac{xx'}{L}\, ,
\label{eq:K_appendix}
\end{equation}
we obtain
\begin{equation}
\Delta t_q^{\rm reg}(E)
=
\frac{1}{2c}\left(\frac{q_\nu}{E}\right)^2
\int_0^{L}\!dx\!\int_0^{L}\!dx'\,
\mathbf{B}_\perp(x)\cdot\mathbf{B}_\perp(x')\,K(x,x')\, ,
\end{equation}
which is Eq.~\eqref{eq:dtq_reg_kernel} of the main text.
\\

The kernel
\begin{equation}
K(x,x')=\min(x,x')-\frac{xx'}{L}
\end{equation}
is the Green's function of the one-dimensional Dirichlet Laplacian on the interval $[0,L]$. Equivalently, after rescaling to the unit interval, it is also identical to the covariance kernel of a Brownian bridge. In the present context, however, its role is purely geometric: it weights the correlated transverse field sampled along the trajectory under fixed source and observer endpoints.

As a simple check, consider a constant transverse field $\mathbf{B}_\perp(x)=\mathbf{B}_0$. Then
\begin{equation}
\int_0^L dx\int_0^L dx'\,K(x,x')
=
\frac{L^3}{12}\, ,
\end{equation}
so that
\begin{equation}
\Delta t_q^{\rm reg}(E)
=
\frac{1}{2c}\left(\frac{q_\nu}{E}\right)^2
B_0^2\,\frac{L^3}{12}
=
\frac{L^3}{24c}\left(\frac{q_\nu B_0}{E}\right)^2\, .
\end{equation}
Thus \(B_{\rm eff}=B_0\), and the general result reduces to the familiar small-angle expression
\begin{equation}
\Delta t \sim \frac{L\,\theta^2}{4c}\, ,
\qquad
\theta \sim \frac{q_\nu B_0 L}{E}\, ,
\end{equation}
up to the expected order-unity geometric factor.

The same kernel formulation also yields the ensemble-averaged delay for a stochastic transverse magnetic field. Let $\mathbf{B}_{\rm stoch,\,\perp}(x)$ denote the transverse component of a stationary, zero-mean stochastic magnetic field sampled along the line-of-sight, and define its two-point correlator by
\begin{equation}
C_\perp(s)
\equiv
\left\langle
\mathbf{B}_{\rm stoch,\,\perp}(x)\cdot \mathbf{B}_{\rm stoch,\,\perp}(x+s)
\right\rangle .
\end{equation}
Averaging the kernel expression over realizations gives
\begin{equation}
\langle \Delta t \rangle
=
\frac{1}{2c}\left(\frac{q_\nu}{E}\right)^2
\int_0^L dx\int_0^L dx'\,
C_\perp(|x-x'|)\,K(x,x')\, .
\label{eq:dt_stoch_start}
\end{equation}
Using the symmetry of the integrand, this may be rewritten as
\begin{multline}
\int_0^L dx\int_0^L dx'\,
C_\perp(|x-x'|)\,K(x,x')
=\\
2\int_0^L ds\,C_\perp(s)\int_0^{L-s}dx\,K(x,x+s)\, .
\end{multline}
Since
\begin{equation}
K(x,x+s)=x-\frac{x(x+s)}{L}
=
\frac{x(L-s-x)}{L}\, ,
\end{equation}
the inner integral is
\begin{equation}
2\int_0^{L-s}dx\,K(x,x+s)
=
\frac{(L-s)^3}{3L}\, .
\end{equation}
Equation~\eqref{eq:dt_stoch_start} therefore becomes
\begin{equation}
\langle \Delta t \rangle
=
\frac{1}{6cL}\left(\frac{q_\nu}{E}\right)^2
\int_0^L ds\,(L-s)^3\,C_\perp(s)\, .
\label{eq:dt_stoch_exact}
\end{equation}

If the propagation distance is much larger than the correlation length, $L\gg\lambda_B$, only separations $s\lesssim \lambda_B\ll L$ contribute appreciably, and one may approximate $(L-s)^3/L\simeq L^2$. This yields
\begin{equation}
\langle \Delta t \rangle
\simeq
\frac{L^2}{6c}\left(\frac{q_\nu}{E}\right)^2
\int_0^\infty ds\,C_\perp(s)\, .
\label{eq:dt_stoch_long}
\end{equation}
Defining the normalized transverse correlation function
\begin{equation}\label{eq:corr_length}
\rho_\perp(s)
=
\frac{C_\perp(s)}{C_\perp(0)}\, ,
\qquad
\lambda_B
=
\int_0^\infty ds\,\rho_\perp(s)\, ,
\end{equation}
one obtains
\begin{equation}
\langle \Delta t \rangle
\simeq
\frac{L^2\,\lambda_B}{6c}
\left(\frac{q_\nu}{E}\right)^2
C_\perp(0)\, .
\label{eq:dt_stoch_C0}
\end{equation}

For an isotropic turbulent field,
\begin{equation}\label{eq:Brms_turb_appendix}
C_\perp(0)=\langle |\mathbf{B}_{\rm turb,\,\perp}|^2\rangle=\frac{2}{3}B_{\rm rms}^2\, ,
\end{equation}
so that
\begin{equation}
\left\langle \Delta t_q^{\rm turb}(E)\right\rangle
=
\frac{L^2\,\lambda_B}{9c}
\left(\frac{q_\nu B_{\rm rms}}{E}\right)^2\, .
\label{eq:dt_turb_appendix}
\end{equation}
This is the origin of the numerical factor \(1/9\) used for the turbulent contribution in the main text.

For an anisotropic component such as the JF12 striated field, the isotropic relation between $C_\perp(0)$ and a full three-dimensional rms field does not apply. In that case, the relevant quantity is the transverse variance itself. Defining $\mathcal{B}_{\rm str}^2 \equiv\left \langle |\mathbf{B}_{{\rm str},\, \perp}|^2\right\rangle_x$

one has
\begin{equation}
C_\perp(0)=\mathcal{B}_{\rm str}^2\, ,
\end{equation}
and therefore
\begin{equation}
\left\langle \Delta t_q^{\rm str}(E)\right\rangle
=
\frac{L^2\,\lambda_B}{6c}
\left(\frac{q_\nu}{E}\right)^2
\mathcal{B}_{\rm str}^2\, .
\label{eq:dt_str_appendix}
\end{equation}
\section{Baseline GMF model and uncertainty treatment}
\label{app:gmf}

Our baseline description of the large-scale regular GMF is the original Jansson--Farrar (JF12) model, which decomposes the coherent field into disk, toroidal-halo, and poloidal ``X-field'' components and was fit to extragalactic rotation measures together with Galactic synchrotron observables~\cite{2012ApJ...757...14J}. In addition to the large-scale regular field, JF12 includes a striated component locally aligned with it and a purely turbulent random component. In our implementation, the regular and striated components are evaluated with in-house routines built directly from the published JF12 parameterization, while the turbulent component is sampled using the \texttt{JF12Field} implementation provided by \texttt{CRPropa}~3.2~\cite{CRPropa:2023app}.

\subsection{Regular component}

To estimate the uncertainty on the regular field contribution to the kernel, we sample the JF12 regular field parameters around their published best fit values using independent Gaussian draws with widths given by the quoted \(1\sigma\) errors. For each sampled parameter set, we recompute the corresponding contribution to the effective field strength \(B_{\rm eff}^2(L,\ell,b)\), and hence to the regular part of the time delay kernel. The uncertainty bands shown in the main text are then constructed from the resulting ensemble of realizations. 

This procedure should be understood as a phenomenological propagation of the published JF12 parameter uncertainties. A full reconstruction of the original JF12 likelihood would require the joint posterior, or equivalently, the full covariance structure, of the fitted parameters, which is not available in the published analysis. The purpose of the present treatment is therefore not to reproduce the original fit statistically, but to quantify the sensitivity of the magnetic delay kernel to the variations of the regular GMF JF12 parameters. Since the benchmark \(1\sigma\) uncertainty bands on the translated charge limits shown in the main text are already at the \(\mathcal{O}(10\%)\) level, residual effects associated with a full covariance-aware treatment are expected to be subleading.

\subsection{Stochastic components}

For the stochastic components, the time delay kernel depends on a transverse field variance together with a correlation length along the line-of-sight given by Eq. \ref{eq:corr_length}. In practice, this integral is evaluated numerically from the transverse correlation function estimated along the line-of-sight.

For the turbulent field, we use the \texttt{CRPropa} realization of the JF12 turbulent component with the regular and striated components switched off. Assuming local isotropy, the full rms field is related to the transverse variance by Eq. \ref{eq:Brms_turb_appendix}. The quantity entering the delay kernel is therefore the product
\begin{equation}
B_{\rm rms}^2\,\lambda_B
=
\frac{3}{2}\,
\left\langle |\mathbf{B}_{\rm turb,\,\perp}|^2 \right\rangle_x
\,\lambda_B .
\label{eq:Brms2lambda_turb_appendix}
\end{equation}
For each line-of-sight and source distance \(L\), we generate multiple independent turbulent realizations, compute the corresponding values of $(B_{\rm rms}^{\rm turb})^2\lambda_B^{\rm turb}$, and use the median together with the 16th and 84th percentiles of the product $(B_{\rm rms}^{\rm turb})^2\lambda_B^{\rm turb}$ to characterize the turbulent contribution.

For the striated component, the isotropic relation of Eq.~\eqref{eq:Brms_turb_appendix} does not apply, since the field is anisotropic by construction: it is locally aligned with the regular field and fluctuates primarily in sign. The relevant quantity is therefore the transverse variance \(\mathcal{B}_{\rm str}^2\) introduced in Appendix~\ref{app:kernel_derivation}. Multiple striated realizations are generated by varying the sign-field realization. For each realization we compute \(\mathcal{B}_{\rm str}^2\), the corresponding correlation length \(\lambda_B^{\rm str}\), and the product \(\mathcal{B}_{\rm str}^2\lambda_B^{\rm str}\), and use the median together with the 16th and 84th percentiles of that product to characterize the corresponding uncertainty.

\bibliography{biblio.bib}

\end{document}